\documentclass[]{article}
\usepackage{geometry}                
\usepackage{graphicx}
\usepackage{amssymb}
\usepackage{enumitem}
\usepackage{epstopdf}
\usepackage{authblk}

\title{A dynamic state-based model of crowds}
\author[1]{Martyn Amos}
\author[2]{Steve Gwynne}
\author[3]{Anne Templeton}

\affil[1]{Northumbria University, UK}
\affil[2]{Movement Strategies, UK}
\affil[3]{University of Edinburgh, UK}

\date{}                                           

\begin{document}
\bibliographystyle{plain}
\maketitle

\begin{abstract}
We consider the problem of categorizing and describing the dynamic properties and behaviours of crowds over time. Previous work has tended to focus on a relatively static ``typology"-based approach, which does not account for the fact that crowds can {\it change}, often quite rapidly. Moreover, the labels attached to crowd behaviours are often subjective and/or value-laden. Here, we present an alternative approach, loosely based on the statechart formalism from computer science. This uses relatively ``agnostic" labels, which means that we do not prescribe the behaviour of an individual, but provide a {\it context} within which an individual might behave. This naturally describes the time-series evolution of a crowd as ``threads" of states, and allows for the dynamic handling of an arbitrary number of ``sub-crowds".
\end{abstract}

\section{Introduction}

As a discipline, crowd science has acknowledged the need to understand the {\it nature} of human collective phenomena before trying to {\it explain} them, and a number of attempts have been made to specify and classify different {\it crowd types and behaviours}. However, these {\it typologies} are often partial, over-fitted to a specific crowd type, or use arbitrary and/or subjective labels for behaviours of complex origin (for example, ``panic"). Moreover, they tend to be relatively inflexible, and do not reflect the fluid nature of crowd behaviour (and how this might influence the crowd's structure and impact over time). For example, a static typology might not capture a situation in which a peaceful demonstration can quickly turn into a riot, or how a physical crowd moving around a shopping mall can suddenly become united into a psychological crowd in response to a shared grievance or an external threat. In this paper, we present an alternative to the typology approach; a {\it dynamic, state-based model of crowds}, structured around an existing {\it assembly-action-dispersal} framework. 

Our model draws on the statechart formalism from computer science. This approach is relatively objective, can capture the dynamic evolution of a crowd over time, and (unlike existing typologies, which are relatively static) allows for the natural description of how sub-groups emerge within a crowd. This new model may be useful for describing the evolution of incidents such as riots or emergencies, but it is equally well-suited to the study of expected, ``normal" crowds. Importantly, our model focuses largely on {\it observable} features of crowds, makes no subjective assumptions about underlying psychological factors or individual motivations, and lends itself to codification in computer simulations.  The model is sufficiently rich to capture macro-level {\it descriptions} of crowds, but may also be easily augmented to provide more fine-grained nuance (for example, by adding probability distributions to state transitions). Our overall goal is to provide a standardised and objective framework for the description of crowds (in terms of both behaviours and properties, and how these dynamically change over time), in the hope that it might find applications in reporting on real crowds and informing future simulation studies. We offer a framework that is amenable to integration with existing modelling approaches, and which is intended to be {\it descriptive} rather than {\it prescriptive}.

The rest of the paper is structured as follows: in Section ~\ref{sec:intro} we describe previous related work; in Section ~\ref{sec:model} we describe the basic model and explain its operation, and in Section ~\ref{sec:casestudy} we illustrate its application with an imagined real-world scenario. We conclude in Section ~\ref{sec:summary} with a consideration of the limitations of our model, as well as suggestions for future work.

\section{Previous work}
\label{sec:intro}

The study of human crowds and their associated dynamics dates back to the late 19th Century \cite{Tarde1903,HobsbawmRev,LeBon1897,Sighele1892} (see also \cite{nye1975origins}), but the field of ``crowd science” (as it would later be referred to) only achieved a degree of methodological rigour around the 1970s \cite{mcphail1971civil,tajfel1979integrative,turnerkillian1972} (see also \cite{mcphail1991myth}). Since then, crowd studies have sought to better understand the complex dynamics of crowds, to predict their behaviour, and to suggest mechanisms by which they may be influenced or even managed (often, but not exclusively, with the aim of trying to ensure the safety of individuals) \cite{challenger2011crowd,drury2009everyone,sime1995crowd,still2014introduction}.

McPhail argues that rigourously capturing the {\it nature} of collective behaviour must preface any attempt to {\it explain} it; ``It is misguided to debate the pros and cons of competing explanations before the phenomena to be explained have first been examined, specified and described" \cite{mcphail1991myth}. In order to motivate what follows, we therefore provide a brief overview of previous attempts to specify and classify crowds and collective behaviour(s). Perhaps the most well-known (in the popular sense) and lyrical codification of crowds comes in Canetti's {\it Crowds and Power} \cite{canetti1962crowds}, in which the  author uses metaphorical images of fire, water and swarms to describe different types of gathering, and provides an extensive catalogue of crowd types and attributes. Although influential, Canetti's work has also been criticised as lacking traditional academic rigour \cite{PhillipsCanetti}.

Prior to Canetti, Blumer  \cite{Blumer1939} initially introduced the notion of` ``acting", ``expressive", ``casual" and ``conventional" crowds,  each instance of a crowd being distinguished by their perceived purpose and dynamics. This framework was modified by Swanson \cite{swanson1953preliminary} (with specific reference to ``acting" crowds) and adopted by Turner and Killian \cite{turnerkillian1972} with the addition of` ``individualistic" and ``solidaristic" crowds. Their typology matrix places ``symbolizing tendencies" (acting and expressive) on one axis, and``coordinating tendencies" (solidaristic and individualistic) on the other, and locates instances of crowds within this two-dimensional space. However, this framework fails to capture attributes or {\it forms} of behaviour, and is not amenable to dynamic reconfiguration. 

McPhail's subsequent work on codifying elementary forms of collective behaviour within gatherings (summarised in \cite{mcphail1991myth}) lists 40 different ``actions" that may be performed by one or more people (e.g., ``booing",  ``bowing"). These have commonly been used by observers to record crowd behaviours, although there are two main arguments against this approach: (a) they are too fine-grained to be able to capture {\it general} attributes of a crowd, and (b) McPhail's work is largely focussed on {\it protest} crowds, and may not be easily modified to capture other forms of gathering.

A relatively influential (and more recent) attempt to characterise crowds is described in \cite{berlonghi1995understanding}, where eleven different crowd types (from ``ambulatory" to ``violent") are specified. However, these types are somewhat arbitrary, and some focus on the ``objective" of the crowd (e.g.``looting") while others describe its physical attributes (e.g., ``dense"), with no distinction made between the two categories. Recent work has focused on the detection of crowd types for the purposes of safety management; a taxonomy constructed to aid the automatic labelling of crowd videos \cite{cheung2016lcrowdv} uses relatively arbitrary definitions (e.g., the difference between ``Assertive" and ``Aggressive" behaviour may not always be clear), and a related paper \cite{wei2020very} also makes a number of assumptions that are perhaps difficult to justify (and relies on the now-discredited theories of Le Bon \cite{LeBon1897}). A design-based approach to crowd categorisation \cite{designerly} uses a diagrammatic methodology to capture the purpose, structure, ``emotional arousal" and movement of crowds, but this provides only a static snapshot (however, the use of a participant {\it timeline}, using different crowd phases, is perhaps instructive for our purposes).

A report prepared for the UK government Cabinet Office \cite{Cabinet2011} highlights the relative scarcity of research into types of crowd (and members of crowds). The authors cite \cite{berlonghi1995understanding}, but emphasise that ``...there is a real need for future research to focus on identifying different types of crowds - categorised according to a broad range of dimensions - along with the characteristics and behaviours they are likely to exhibit. Ultimately, this should assist event planners and managers with their preparation for, and management of, a particular crowd event with a particular type of crowd comprised of particular types of crowd member." The report proposes a crowd typology, which includes categorisation dimensions such as  crowd purpose, duration, start time, individual locations,  event atmosphere, and so on. Each category is then subdivided into subclasses (for example, ``purpose" is divided into ``sport/entertainment", ``travel", ``religious meeting/political demo", ``spontaneous", and ``mixed"). The authors state that such a typology ``...could be used as a framework to help think through the numerous characteristic and likely behaviours which different types of crowd may have and, consequently, should help to guide event preparation."

While this typology is an advance on what went before, it still imposes a ``top down" structure on crowd characteristics, and it is difficult to see how it easily admits the possibility of {\it emergent} (or unexpected) behaviours. Typologies such as this, arranged into a hierarchical, tree-like structure, view the crowd as essentially unchanging, without understanding where behaviours stem from. However, in reality, the behaviour of a crowd is inherently {\it dynamic} and {\it complex}, and is influenced by a number of factors (including, but not limited to, heterogeneity of its composition, size and density, physical location and space, emotions, communication, group dynamics, and other external and contextual factors) \cite{haghani2023roadmap}. 

We therefore believe that a dynamic, {\it transitional} model is needed, which captures the changing behaviour and attributes of a crowd {\it over time}. Moreover, existing typologies still use relatively arbitrary labels for crowd types (e.g., ``Religious meeting/political demo"), which may not capture all possible purposes. In what follows, we describe a new, more general model of crowds that captures a wide range of different gatherings, allows for dynamic crowd reconfiguration, is easily extensible, and offers the possibility of more formal encoding at a later date. This model is particularly well-suited to capturing the {\it evolution} of a crowd over time, and imposes relatively few arbitrary or subjective labels on participant behaviour.

Fundamentally, what we seek is a way of {\it conceptualizing} the crowd that allows for dynamic reconfiguration and the incorporation of important concepts such as social identity \cite{seitz2017parsimony,templeton2020modeling}, and which supports a move away from the ``masses" model of crowds to embrace a more nuanced group-based approach \cite{templeton2015mindless}. Our model is partly inspired by the early work of Sime, who presented his {\it decomposition diagram} for domestic fires in \cite{sime84}. This work codified multiple verbal/written incident reports into a general descriptive framework for ``act sequence analysis". Using this model, it is possible to describe different individual sequences of behaviours in a consistent manner, as well as capturing common patterns of behaviour.

\section{Our dynamic state-based model}
\label{sec:model}

Our model is sufficiently general that it can be applied to both  {\it physical} \cite{degond2013hierarchy,moussaid2009collective} and {\it psychological crowds} \cite{drury2009everyone,reicher2001psychology,templeton2018walking}. The physical  crowd is made up of people who simply happen to be in the same place at the same time, while the psychological crowd exhibits some form of common social identity (a sense of connectivity or feeling of being part of the same group) \cite{hornsey2008social}. With that in mind, we use the sub-definition of ``crowd" given in \cite{adrian2019glossary}:

\begin{quote}
``A (typically large) number of people in one place at the same time. It is possible that a physical crowd contains one or more psychological crowds (e.g. football fans in a transport hub with commuters)."
\end{quote}

We assume a four-phase model of crowds, inspired by the three-phase ``convergence-task-divergence" life cycle framework \cite{Adang2016}). The {\it Assembly} phase describes the initial formation of a crowd, and the {\it Mode} phase specifies a crowd's rationale (i.e., reason for assembling). The {\it Structure} phase captures a crowd's physical configuration, and the {\it Dispersal} phase describes the ``break up" of the crowd. This separation into four phases avoids issues such as the conflation of rationale and structure, whilst giving the flexibility to capture a full range of interactions and dynamic behaviours.

\begin{figure}[]
\begin{centering}
\includegraphics[scale=0.60]{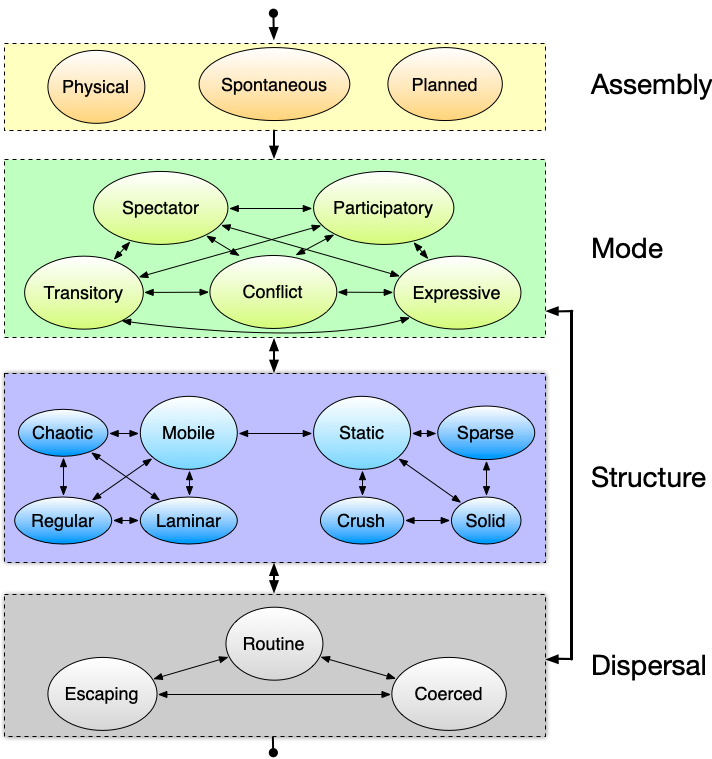}
\caption{Dynamic state-based crowd model, showing Assembly (yellow), Mode (green), Structure (blue) and Dispersal (grey) phases.}
\label{fig:model}
\end{centering}
\end{figure}

Our general model is depicted in Figure ~\ref{fig:model}. We use a loose interpretation of the {\it statechart} formalism from computer science \cite{harel1987statecharts}, whereby a system (that is, the crowd itself) may exist in one of a finite number of {\it states} (denoted by coloured ovals), which are clustered across the four phases. In what follows, we textually denote states as ``Phase:State" (e.g., Structure:Chaotic).

The model is initialised by moving into the Assembly phase, and then moves between connected states (this is called making a {\it transition}), corresponding to some change in the crowd. Transitions may be triggered by observable (or, in the case where the model is being used as part of a simulation, algorithmically-generated) changes in the crowd's behaviour, density, and so on, but they may also be {\it externally} generated by a trigger such as police action.

Most transitions in our representation are bidirectional (e.g., a crowd could move from Mode: Spectator to Mode:Participatory, and then back to Mode:Spectator), but transitions may also be one-way (e.g., after the Assembly phase has completed, or when the crowd has completely dispersed and enters the so-called ``terminal" state at the bottom of the diagram). 

A transition may also be made from a state in one phase (i.e., cluster) to any state in a {\it different} phase (for example, from Mode:Spectator directly to Dispersal:Escaping), as shown by the bidirectional lines connecting phases. In this way, appropriate levels of abstraction may be implemented (for example, if detailed information on the crowd's structure is not available), and this also avoids the need for the model to undergo an artificial or potentially incorrect series of transitions in order to move from one phase to another (as would be the case if the transition from one phase to another had to be made through a specific state). While this capability perhaps loosens the connection with the statechart formalism, we gain significant benefits in terms of flexibility and ease of description.

As in \cite{sime84}, within our formalism the evolution of a crowd is captured as a subset of states and the transitions between them (which we call a ``thread"), which essentially describes a time series of events and crowd attributes. Importantly, this statechart formalism, whereby all states are encapsulated into a single data structure, allows for the easy construction of sub-groups/clusters within crowds. If, for example, a crowd is considered to be in a uniform state, then a single ``thread" of the statechart is sufficient to capture that crowd. However, if part of the crowd begins to exhibit different behaviours or properties (e.g., a region of higher crowd density, or a more expressive character), then the statechart may branch off at that point to create a new threads (series of states) which is seperately maintained, and that region of the crowd associated (perhaps temporarily) with the new thread (which has its own independent existence). As we will later demonstrate, the diagrammatic representation of this ``threading" may include supplementary information (such as timestamps or details of provoking incidents) in order to track the evolution of the crowd in an intuitive fashion. In this way, an arbitrary number of clusters/threads may be captured simultaneously within a single diverse crowd, which is one of the main benefits of our approach.

The model is ``transparent" in the sense that the current state may be inspected at any time from ``outside" the model, and we note that it allows for a natural ``time series" representation of a crowd. By easily allowing for transitions to be made between both states and clusters, we facilitate rapid {\it dynamic restructuring} of a crowd's description, which encompasses its assembly, its overall rationale, aspects of its embodied physical ``architecture", and its eventual dispersal. This allows a crowd (and and subsequent sub-crowds) to move between different modes, physical structures and ways of dispersal, thus preventing over-generalisation and ``lock in" to a static state (as implied by most typologies). We now describe, in detail, the states clustered within each phase.

\subsection{Assembly phase}

The assembly phase is depicted by the yellow states; we necessarily select an arbitrary ``start point" for consideration of the crowd event. After that time, a crowd may assemble in one of three basic ways:

\begin{itemize}

\item Physical, where individuals congregate or move through a specific area, but {\it without common purpose} (e.g., a shopping street or transport hub). Such assemblies are also referred to as ``casual" crowds, as there is no real  shared identity present.

\item Spontaneous, where individuals form a crowd (or sub-crowd) {\it in response to a specific ``trigger" event} (e.g., to watch a street performance, to act as bystanders at an incident (emergency or otherwise), or as a result of some external intervention).

\item Planned, where individuals gather for a {\it specific purpose}  (e.g., a concert, sporting event). Crowds assembling in such a way are sometimes referred to as ``conventional" (if the event is spectator-based, like a concert) or ``expressive" (if the event is largely characterised by the expression of opinion, as in a political rally). This leads us naturally onto a consideration of the {\it purpose} of a crowd's existence, which we detail in the next phase.
\end{itemize}

\newpage
\subsection{Mode phase}

Once the crowd has assembled, it may move into the main Mode phase (depicted in Figure ~\ref{fig:model} as the green cluster of states). In this phase, we have states representing the overall crowd {\it rationale} (i.e., purpose for existing). The four main categories of crowd rationale we identify are as follows:

\begin{itemize}

\item Spectator: the primary purpose of such crowds is to {\it watch or observe} a specific event (e.g., concert/cinema audience). Of course, we admit the possibility of audience participation in the event (e.g., applause, singing along), but this is not the {\it primary} rationale for the crowd's existence.

\item Participatory: these crowds display {\it active involvement} from individuals (e.g., a religious procession), without necessarily being characterised by expressions of opinion (see below).

\item Transitory: these crowds are generally characterised by {\it movement of individuals}, such as may be found in a busy street, or commuters moving through a transportation hub. 

\item Conflict: these crowds are characterised by {\it hostility} or violence beyond the level of ``expression", which may be expressed internally (e.g., a fight between members of the crowd), towards the environment (e.g., the destruction of property) or in relation to another group (e.g., police).

\item Expressive: such crowds largely exist for the purposes of {\it expression} (e.g., chanting at a political rally, or taking part in a celebration). Of course, there may be significant intersection with the ``spectator" or ``participatory" states. 
\end{itemize}

\subsection{Structure phase}

Once a crowd is assembled and its mode is established, we might then consider its physical configuration, represented by the blue Structure states. Importantly, we {\it separate out} Mode and Structure states, which addresses a significant flaw in previous typology-based approaches. Berlonghi \cite{berlonghi1995understanding}, for example, conflates the {\it purpose} of a crowd with its {\it physical characteristics}: examples of different crowd types within this typology include ``spectator" and ``dense/suffocating", implying that these crowd types are somehow disjoint. In fact, as many incidents have shown (e.g., fatal crowd crushes that occurred at Hillsborough \cite{nicholson1995investigation} and the Love Parade \cite{helbing2012crowd}), crowds should be able to simultaneously exhibit {\it more than one crowd type} under the typology model (for example, a dense/suffocating spectator crowd), and it is not immediately clear how this may be represented using a static typology. However, our model easily admits this by permitting transitions between states (and, perhaps, the persistent recording of phase states) in a natural manner.

As we record the evolution of a crowd as a sequence of state transitions, once a crowd has entered a Mode state, we assume that it remains in that ``mode of thought", unless it explicitly enters another Mode state (thus changing its purpose). In this way, for example, a Mode:Transitory crowd can transition into Mode:Participatory, but also begin in Mode:Participatory and then cycle through a number of physical Structure states (whilst implicitly retaining the Participatory mode); as we discuss in a later Section, this retention may be explicitly codified using persistent ``tags" to serve as a form of memory.

We identify two main physical configurations that a crowd may take, each of which has three sub-configurations:

\begin{itemize}
\item Mobile crowds contain individuals in motion. This movement may be Chaotic (little, if any, coordination between individuals; movement is characterised by turbulence), Regular (essentially similar to that generated by standard movement models without the existence of significant internal structure), or Laminar (layers or ``streams" exist) \cite{helbing2007dynamics}.

\item Static crowds involve little or no net movement (of course, individuals are free to enter and leave, if they are able). Such crowds may be Sparse (identifiable gaps between individuals), Solid (few gaps exist between individuals) or in a Crush state (no space between individuals, individuals are very tightly packed), each signifier corresponding to a specific density (low, medium and high, respectively) \cite{still2014introduction}. Of course, any implementation of the model may apply its own density thresholds for each of these states.
\end{itemize}

\subsection{Dispersal phase}

From any state in the Structure phase, the crowd may move into one of the three grey Dispersal states. These are entered when the crowd begins to dissipate:

\begin{itemize}
\item Routine: the crowd disperses naturally in a ``normal" fashion (this includes being ``merged" into another crowd).
\item Escaping: the crowd willingly leaves an area in response to a threat or other emergency, perhaps (but certainly not always) in response to information given or requests issued by the authorities. This state may include situations in which members of the crowd are {\it rescued} or similarly removed (e.g., from a crush).
\item Coerced: the crowd is {\it forcibly} dispersed (against the will of its members) by the authorities (or even, perhaps, by another group).
\end{itemize}

Importantly, a crowd in this phase may move {\it back} into either the Structure or the Mode phase. This allows for the possibility, for example, of a Structure:Solid crowd dispersing in a Routine fashion, at which point an incident occurs (e.g., a blocked stairwell), causing the crowd to move back into a rapid sequence of transitions from Structure:Static to Structure:Crush, after which point we move into a Dispersal:Escaping state. Alternatively, a Routine dispersal could directly turn into (for example) a Coerced dispersal, and so on. This further illustrates the flexibility afforded by a dynamic typology such as ours.

\subsection{Possible extensions}

Although we do not explicitly use them here, in order to control the probability distribution of a crowd moving between states, it is sometimes desirable to attach indicative {\it weights} to transitions. In this way, the likelihood of a crowd making a transition from one state to another may be mathematically encoded. This is particularly relevant if we are using our model to generate plausible crowd evolution sequences as part of a modelling exercise.  

So, for example, if observations show that crowds generally move from one state to another 50\% of the time, that particular transition might be given a value of 0.5. If all transitions are weighted in this way, then realistic crowd state sequences may be easily simulated by simply performing a probabilistic ``walk" on the network, using the transition weights to move between states, and thus provide a time series of crowd behaviour/properties. 

In order to precisely define these transitions, it is important to have {\it metrics} with which to assess the state of the crowd. It is relatively difficult to objectively measure the difference between Mode:Participatory and Mode:Expressive, although some attempts have been made to quantify expression in terms of crowd noise and participation \cite{hutton2011assessing}. Crowd movement and density is easier to objectively measure, and metrics based on information theory and image analysis have been used successfully to detect clusters of behaviour  \cite{conigliaro2013viewing} and transitions from laminar to chaotic flow, which often signals the onset of dangerous crush conditions \cite{harding2011mutual}.

We may also build a form of ``memory" into the model in the form of a persistent ``tag" for each phase, which stores the last state that was entered within that phase. So, for example, a crowd may assemble in the Assembly:Planned state, and then transition naturally into the Mode:Spectator state. However, on leaving the Assembly phase, the yellow tag stores the fact that the crowd was last in the Assembly:Planned state. Of course, tags may also be overwritten if we move back into a previously-departed phase (for example, if a political rally crowd moves from Mode:Spectator (while listening to a speaker) to Structure:Mobile (beginning to move as part of a march) and then back to Mode:Expressive to reflect the fact that it is now largely characterised by chanting and so on, in this case, the Mode tag would be overwritten from Spectator to Expressive.) 

\section{Example application}
\label{sec:casestudy}

In this Section we illustrate the application of our model by way of an imagined scenario - a city-centre rally taking place in a square (inspired, in part, by \cite{drury1999intergroup}). We first describe the evolution of the event, using the traditional form of text-based narrative and diagrams, and then show how this form of description might be mapped onto our model. By illustrating the mapping, we show how such informal descriptions may be {\it codified} in a consistent manner, as well as potentially allowing for integration with algorithmic analysis. In Figure ~\ref{fig:rally} we show a series of events, as follows:

\begin{figure}[]
\begin{centering}
\includegraphics[scale=0.60]{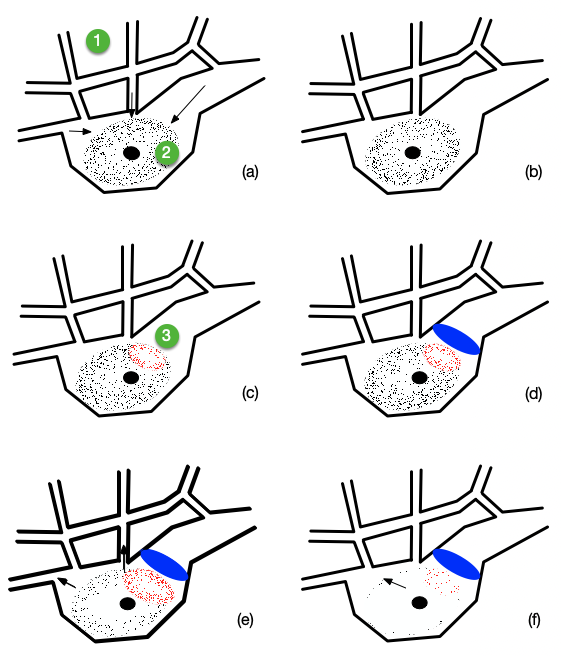}
\caption{Evolution of an imagined crowd event. (a) Crowd gathers in square, with two main groups - (1) the mobile ``streams" and (2) those already assembled. (b) The audience has assembled. (c) A sub-group (3) begins to demonstrate. (d) The police respond with a cordon. (e) This creates conflict, and some members of the crowd leave in response. (f) The remainder of the crowd is dispersed by the police.}
\label{fig:rally}
\end{centering}
\end{figure}

\begin{enumerate}[label=\alph*]
\item The crowd is assembling in the square. We identify two main clusters; (1) is made up of streams of people heading for the square, and (2) is comprised of the people who are waiting for the rally to begin.
\item The crowd has largely assembled, and is listening to a number of speakers.
\item A section of the crowd begins to loudly chant anti-government slogans. This distinct new group is cluster (3).
\item In response to this, the police deploy a cordon of officers to contain and/or disperse the crowd, which creates a state of conflict with the protestors (possibly also increasing the size of the protesting group).
\item The protesting group is now engaged in violent confrontation with the police, while other members of the crowd are escaping the scene.
\item The protestors are eventually dispersed by the police.
\end{enumerate}

\begin{figure}[]
\begin{centering}
\includegraphics[scale=0.40]{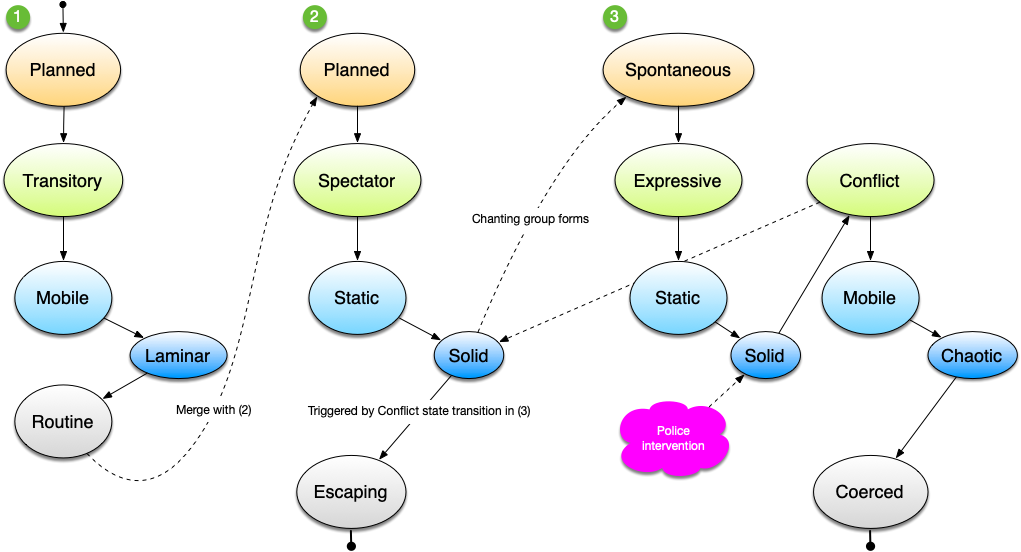}
\caption{Mapping of the event onto our model. Threads (1-3) represent the sub-crowds identified in our analysis.}
\label{fig:demomodel}
\end{centering}
\end{figure}

In Figure ~\ref{fig:demomodel} we show how this description might be coded within our model. We begin with an Assembly:Planned crowd (labelled (1)), which is Mode:Transitory, Structure:Mobile and Structure:Laminar (to represent streams of people converging on a central point). When enough people have accumulated at the meeting point and the rally has begun, we create crowd (2), which is Assembly:Planned, Mode:Spectator (listening to speakers), Structure:Static, Structure:Solid. Eventually, the protesting group (3) emerges, which is Assembly:Spontaneous, Mode:Expressive (but still Structure:Static, Structure:Solid). An external event is triggered (the police intervention, denoted by the magenta cloud), acting on this group, which forces a transition into Mode:Conflict. The transition into the Conflict state also triggers a transition in crowd (2), which moves into Dispersal:Escaping (to represent people moving away from the conflict), and the thread terminates. The protesting group (3) adopts Structure:Mobile and Structure:Chaotic, to represent engagement with the police. Eventually, this group moves into Dispersal:Coerced state, and the thread terminates.

\section{Summary and further work}
\label{sec:summary}

In this paper we have described a dynamic state-based model of crowds that is capable of dealing with changes in assembly, rationale, physical structure and dispersal over time. It is sufficiently general to encompass a wide variety of crowds, and is particularly well-suited to handling the formation of sub-groups.
Importantly, we present this simply as a starting point, in a descriptive (rather than prescriptive) fashion. The model may be easily augmented or modified to suit a particular purpose - what we propose here is a general conceptual architecture for thinking about the evolution of crowds over time, rather than a rigid typology. We acknowledge a number of limitations of our model. We still make some relatively subjective decompositions (e.g., in terms of the physical states, which may not capture specific structures in sufficient detail). The issue of {\it granularity} is still relatively fuzzy (e.g., at what point do we decide to create a new thread to represent the emergence of a new group, area of differential density, etc?). We also note that the transitions between discrete states are not always sharp; rather, crowds/groups may exist on a {\it continuum} between states. However, we believe that this model offers a new framework for thinking about and representing the types of dynamic macroscopic outcomes that emerge from the interactions between individuals and their environment \cite{gwynne2018model}. \\

The need for a dynamic model is particularly important with the growth in reliance on simulation tools within crowd science and the practice of crowd management \cite{haghani2021knowledge}. As noted in \cite{Cabinet2011}, ``The next generation of simulation tools should be able to accommodate multiple environments and multiple crowd events in the same simulation model, rather than solely modelling one particular event in one particular environment." This is precisely what our model offers, and future work will focus on both investigating how it might be incorporated into crowd simulation platforms, and using it to analyse historical footage and accounts of crowd-incidents, in order to compile a set of case studies. We believe that this will be particularly valuable in terms of capturing and codifying complex event narratives (perhaps, in future work, using an algorithmic or AI-based approach), where multiple events occur in parallel, external factors are significant, and the emergence and behaviour of sub-crowds is highly dynamic (for example, both the Hillsborough and Love Parade disasters).

\section*{Acknowledgements}

The authors thank Otto Adang, John Drury and Keith Still for helpful discussions and useful feedback. AT is supported by a UKRI Future Leaders Fellowship (MR/V023748/1).

\bibliography{crowd}

\end{document}